\RequirePackage[2020/08/10]{latexrelease}
\documentclass[aps,pra,reprint,superscriptaddress,longbibliography]{revtex4-2}

\usepackage{physics}
\usepackage{amsmath}
\usepackage{amssymb}
\usepackage{graphicx}
\usepackage{float}
\usepackage{mathrsfs}
\usepackage{mathtools}
\usepackage{extarrows}
\usepackage[margin=0.7in]{geometry}
\graphicspath{{PNG/}}
\usepackage{xcolor}

\begin{document}
	
	\title{Switching of post quench reflection asymmetry in an embedded non-Hermitian Su-Schrieffer-Heeger system}
	
	\author{A.~ Ghosh}
	
	\affiliation{School of Physics, University of Melbourne, Melbourne, 3010, Australia}
	
	\affiliation{Indian Institute of Technology Kharagpur, Kharagpur 721302, West Bengal, India}

	\author{A.~M.~Martin}
	
	\affiliation{School of Physics, University of Melbourne, Melbourne, 3010, Australia}

	\date{\today}

	\begin{abstract}
	A quench in a Su-Schrieffer-Heeger lattice across the topological boundary initialized with an edge state leads to transport across the chain. We consider such a quench in an effective model in which non-Hermitian components are embedded in an SSH lattice. We find that the transport arising as a result of quench is asymmetric in the sense that there is imbalance in reflection in transport from left and right and this imbalance switches from higher right reflection to higher left reflection as the configuration to which the system is quenched varies in parameter space. We discuss the switching as emergence from the underlying phenomenon of a partial reorganization of bulk states in terms of localization and energy, the intricacies of which depends upon the configuration of the system and the symmetries present.

	\end{abstract}
	
	\maketitle

	\section{Introduction \label{sec:1}}
The field of non-Hermitian physics has garnered significant attention, particularly after pioneering studies demonstrated that non-Hermitian systems with PT symmetry can exhibit real energy spectra, thus representing physical systems \cite{bender1998real, bender2005introduction, bender2007making}. This realization quickly extended to photonic systems, which, despite being classical, emulate the mathematical structure of the Schrödinger equation, with the propagation direction serving as an analogue to time \cite{klaiman2008visualization, ashida2020non}. As a result, the field of photonics has experienced substantial theoretical and experimental advancements, particularly in the exploration of exotic phenomena involving PT-symmetric Hamiltonians and exceptional points \cite{ makris2008beam, el2007theory, mostafazadeh2009spectral, guo2009observation}. The rich physics of non-Hermitian systems has also been explored in other classical settings, such as mechanical \cite{ashida2020non, yoshida2019exceptional}, acoustic \cite{fleury2015invisible}, and electrical systems \cite{schindler2011experimental}.
In quantum systems, the use of non-Hermiticity dates back to use in nuclear physics for explaining radioactive decay \cite{gamow1963quantum} and Feshbach resonances \cite{feshbach1958unified, feshbach1962unified}, which essentially described open quantum systems through a projection formalism. More recently, non-Hermitian quantum mechanics has found applications in a broader range of contexts, including many-body systems with losses \cite{ashida2020non}. 
The field of topological physics has flourished, marked by remarkable progress in both theoretical frameworks and experimental realizations \cite{asboth2016short, bernevig2013topological, hasan2010colloquium, zeuner2015observation}. This convergence of research has sparked an exciting new frontier where the principles of non-Hermitian physics are extended into the realm of topological phenomena, revealing intriguing effects such as the non-Hermitian skin effect and unconventional bulk-boundary correspondence \cite{bergholtz2021exceptional}. 
 The Su-Schrieffer-Heeger (SSH) model \cite{su1979solitons, asboth2016short}, a fundamental 1D system known for hosting topological edge states, has recently been investigated in various non-Hermitian settings \cite{lieu2018topological, halder2022properties}. Extensions of this model have demonstrated intriguing dynamical behavior, particularly in the variation of transport velocities along paths that share identical set of initial and final topological invariants \cite{ghosh2023quench}.\\
  In this paper, we study the transport that arises as a result of a quench in an effective model, which is an SSH lattice embedded with non-Hermitian components. We find asymmetry in the transport, i.e., an imbalance in the reflection from the non-Hermitian component for the transport coming from the left and the right. The polarity of the transport imbalance depends heavily on the location of the quench in parameter space, meaning this polarity switches as the final quench state traverses the parameter space. This switching in the polarity of the imbalance emerges as a consequence of the reshuffling of bulk-states in the system with respect to parameter space when non-Hermitian components are present, the intricacies of which are dependent on the details of the system under consideration which includes symmetry. 
This paper is organized as follows; Section II, provides the intuition behind studying quench in current system by briefly demonstrating that a quench across topological regimes in SSH lattice leads to transport that forms light cones. This leads to the introduction of the effective model in Section III, where, we argue the condition for the key topological features to be preserved. We also demonstrate some important features such as the position of critical points for various configurations. In Section IV(part A), we demonstrate the phenomenon of switch in polarity of transport asymmetry depending on the position in parameter space to which the system is quenched using light cones, bipartite probability, etc. Additionally, we highlight these characteristics of transport across a range of configurations. In Part B of this Section(IV), we discuss how this switching of polarity in reflection, with respect to the final quench configuration, emerges as a result of the re-organization of bulk states in terms of energy and localization. A summary is provided in Section 5.

	\section{Quench in a SSH system \label{sec:2}}
As a starting point, we consider a quench in the SSH system, which is a simple tight binding 1D model that can host topologically robust edge states \cite{asboth2016short, su1979solitons} and is constructed by the staggered hoping amplitudes $ v $ and $ w $. The Hamiltonian is given by,
	\begin{equation}\label{eqn:2}
		\hat{H}= v \sum_{n=1}^{N} \ket{A,n}\bra{B,n} + w \sum_{n=1}^{N-1} \ket{A,n}\bra{B,n+1}+h.c,
	\end{equation}
	 where,  $ v $ is the intracellular hopping, and $ w $ is the intercellular hopping.
 The system can support topological edge states for \( v/w < 1 \), and these states are highly localized at the edges. As \( v/w \) approaches 1, the edge states gradually delocalize, the band gap closes at \( v/w = 1 \), and beyond this point (\( v/w > 1 \)), the band gap reopens and the edge states ceases to exist.
	Let’s examine the system’s dynamics as it is quenched from an initial configuration “\(i\)”, with the initialized edge state \(\ket{i}\)(right/left), to a final configuration “\(f\)”, a regime where \(v/w\) exceeds 1, i.e., the edge-states are absent.
	The evolution is governed by the expression $ \ket{\psi(t)} = e^{-i \hat{H} t}\ket{i} $ (in units where $ \hbar=1 $), which along with the completeness in terms of the states of the final Hamiltonian, $ \sum_{f}\ket{f}\bra{f} = I $, gives,
	\begin{equation}
		\ket{\psi(t)} = \sum_{f}\bra{f}\ket{i}e^{-i E_{f} t}\ket{f},
	\end{equation}
	where $ E_{f} $ is the energy of the final state $ \ket{f} $. Equation 2 leads to the transport\cite{ghosh2023quench}, which we have demonstrated in Figure 1 as a light cone representing the probability density $ \rho = \bra{\psi_{n}}\ket{\psi_{n}} $ on each site $ n $ with respect to time $ t $. It may be observed that the probability density is transported from the initial localization position to the other end, eventually returning to original position after reflection from the other end. 
	\begin{figure}[H]
		\centering
		\includegraphics[width=1.0\linewidth]{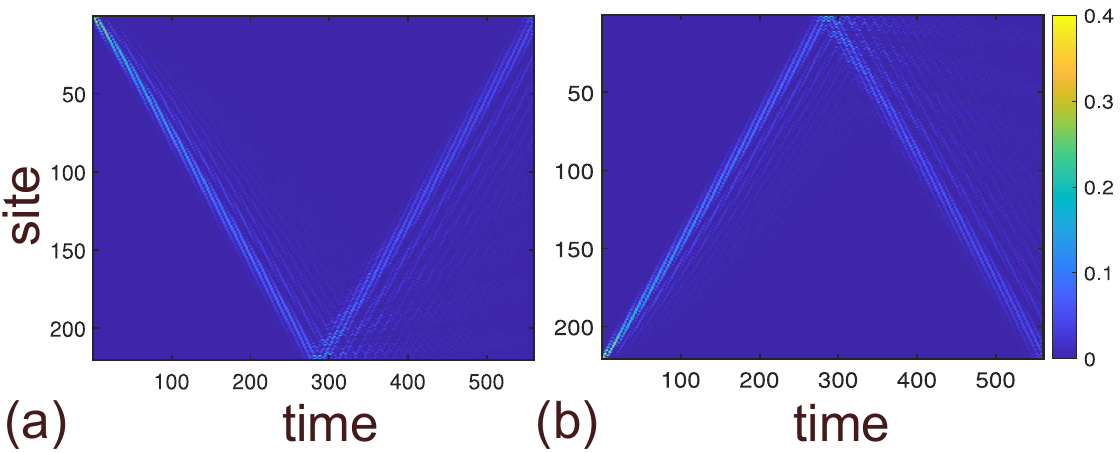}
		\caption{Light cone diagram denoting the transport of probability density after a quench to a configuration of $ v/w = 1.5 $, where, the initial configuration hosts topological edge states with $ v/w = 0.25 $, and the initial state is chosen as edge sate localized on (a) left edge and (b) right edge.}
		\label{fig:figure-1}
	\end{figure}
This raises intriguing questions: how might the introduction of non-Hermitian elements affect the post-quench transport dynamics? What characteristics will emerge, and how will they depend on the specific configuration? To explore this, we examine an effective model, which will be discussed in the next Section.

	\section{Introduction to the system: Embeded non-Hermitian SSH}
	Keeping in mind the transport from Section II, in order to explore the rich physics associated with non- Hermitian PT-symmetric systems, here we will incorporate some non-Hermitian components by considering a hybrid system that consists of non-Hermitian components embedded inside an SSH system. The idea is to first study the rich dynamics and then understand its emergence from the unique underlying static physical properties. The Hamiltonian for such a system will have the form,
	\begin{align}\label{eqn:2}
		\hat{H}&= v \sum_{n=1}^{N} \ket{A,n}\bra{B,n} + w \sum_{n=1}^{N-1} \ket{A,n}\bra{B,n+1}+H.c \\
		&+ \sum_{n=n1}^{n=n2} u\ket{A,n}\bra{A,n} + u^{*}\ket{B,n}\bra{B,n}\nonumber.
	\end{align}
	Here the $ u=U_{Re} -i U_{Im} $ and $ u^{*}=U_{Re}+ i U_{Im} $ are the alternating on-site complex potentials and the non-Hermitian region extends from site $ n_{1} $ to site $ n_{2} $. The schematics of the system is demonstrated in Figure 2(a), whereas Figure 2(b) presents the geometric intuition that the dimerized case ($v/w=0$) at least should host edge states. if the number of non-Hermitian components are small, for a wider range of $ v/w $,  it is straightforward to see that this system will host edge states by the fact that this system may be considered as a perturbation to the pure SSH system, i.e., $ \hat{H} = \hat{H}_{SSH} + \hat{H}^{'} $, for $ \hat{H}^{'} = \sum_{n=n1}^{n=n2} u\ket{A,n}\bra{A,n} + u^{*}\ket{B,n}\bra{B,n} $. The correction to the zero energy \( \bra{\psi_{edge}}\hat{H}^{'}\ket{\psi_{edge}} \), given a small number of PT-symmetric pairs as impurities or when the matrix elements are centered in the perturbation matrix \( \hat{H}^{'} \), should be nearly zero because the zero-energy states are localized at the edges. 
	\begin{figure}
		\centering
		\includegraphics[width=1.0\linewidth]{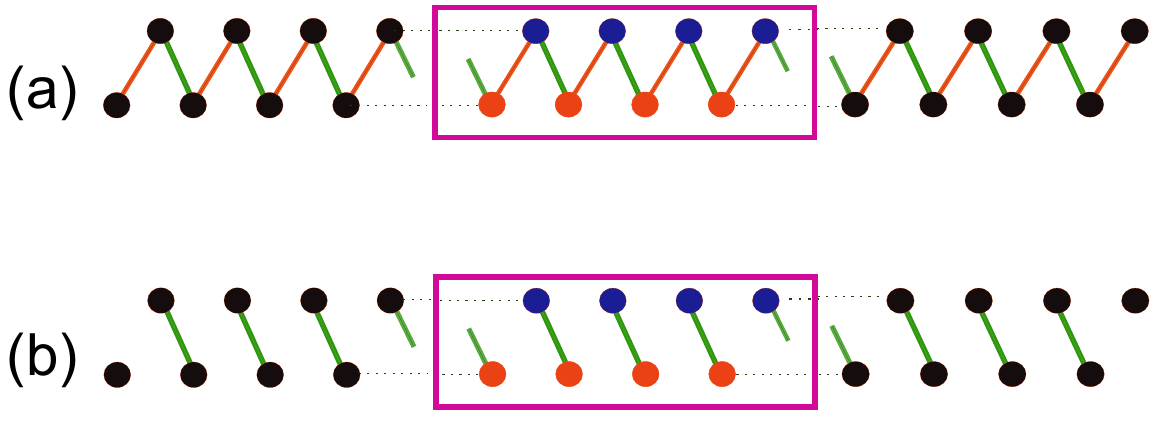}
		\caption{(a) The cartoon representation of the effective model where the circular points denote the lattice sites and the orange and green lines indicate the intra-cellular($v$) and inter-cellular(w) hopping, respectively. The colors on the circular points inside the boxed region represents the superimposed on-site complex potential pairs $ U $(orange) and $U^{*}$. (b) Shows the geometric intuition that under the extreme dimerized case($ v=0 $), there will be a isolated zero energy state at the edges. }
		\label{fig:figure-2}
	\end{figure}
	\begin{figure}
		\centering
		\includegraphics[width=1.0\linewidth]{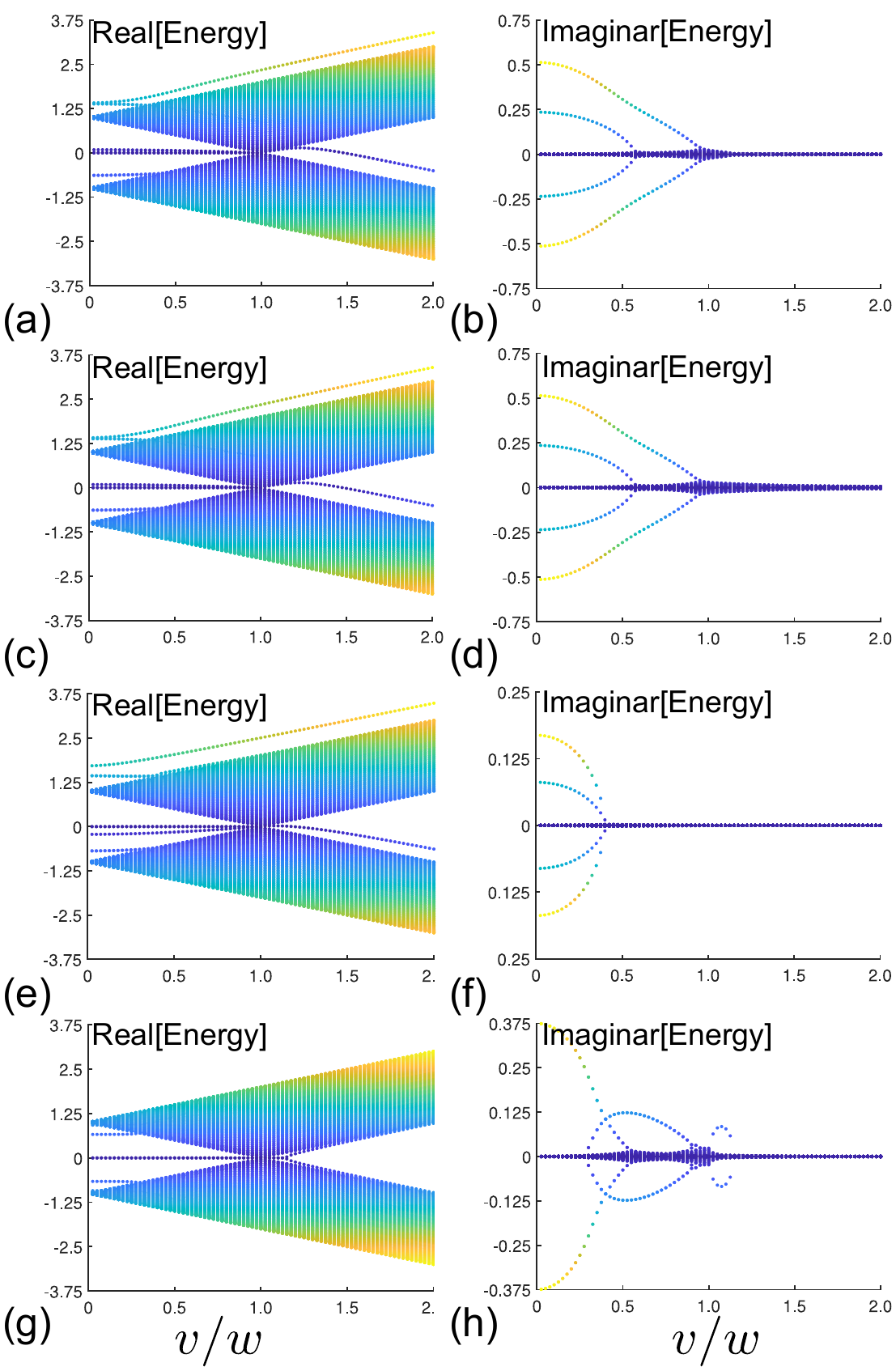}
		\caption{The real (left column) and imaginary (right column) components of energy in parameter space ($v/w$) for different configurations (rows), where the non-Hermitian terms ($U_{Re}/w \mp U_{Im}/w$) are applied to a Su-Schrieffer-Heeger (SSH) lattice of 220 sites (110 unit cells).The color coding scales how far the absolute value of the energy is from zero. In panels (a) and (b), non-Hermitian pairs (0.75, 0.75) are placed on sites 109-112, maintaining PT symmetry. In (b), the critical point in parameter space, where the energy levels coalesce, occurs near the closure of the gap in real energy. Panels (c) and (d) depict the same potentials as (a) and (b), but positioned on sites 107-110, breaking PT symmetry. In (d), the imaginary energy trail is thicker compared to (b). Panels (e) and (f) show a PT-symmetric configuration on sites 109-112, with potentials having a lower imaginary component (0.75, 0.25). In (f), the critical point appears much earlier than the gap closure in real energy (e). Panels (g) and (h) feature purely imaginary non-Hermitian components (0, 0.75) placed on sites 109-112, resulting in a symmetric energy distribution (real and imaginary) around zero.
		}
		\label{fig:figure3}
	\end{figure}
	Now, if the number of non-Hermitian PT symmetric pairs is increased gradually, more number of bulk states will shift, keeping the zero energy states fixed and as a result of this selective shift there will be an overlap of the bulk states and zero energy edge states from some particular value of $ v/w $, whence onwards the localized behviour of the zero energy states will be lost. However, to meet our objective, the initial edge state must be localized. This can be achieved by selecting few PT-symmetric pairs as impurities to minimize the overlap of the bulk states with the zero-energy edge state, and/or by choosing a regime of \(v/w\) for the initial Hamiltonian where no overlap occurs.  The further details of the energy spectrum depend upon the details of the non-Hermitian components, such as their values, number, and how they are placed in the system. Figure 3 provides some examples of different configurations. In all of these examples, the zero energy states remain isolated from the bulk states, and hence, the edge states exist till $ v/w=1 $. For non-Hermitian systems, a point in parameter space($ v/w $ in this case) exists where the complex energies merge \cite{bender2005introduction, bender1998real}. The location of these points may or may not converge with the position of the phase transition, where the edge state loses its existence. For example, in Figure 3((a),(b)), in which the non-hermitian pairs($(U_{Re}/w, \mp i U_{Im}/w) = (0.75, \mp i 0.75)$) are placed on sites $ 109-112 $, the system becomes PT-symmetric, and the exceptional point lies close to the point of topological phase transition. Placing the same pairs slightly shifted from the central part(107-110) breaks the PT-symmetry (Figure 3(c),3(d)), and the imaginary values do not merge completely, i.e., do not reach zero, which may be observed by comparing the thicker trail of imaginary values of Figure 3(d) with respect to Figure 3(b) following $ v/w =1 $(also will be discussed in further detail in Section IV(b)). For the case when non-Hermitian components have the value $(U_{Re}/w, \mp i U_{Im}/w) = (0.75, \mp i 0.25)$, the position of the exceptional point is distinctly different from that of the position of the topological phase transition, as shown in Figure 3((e),(f)). Making the real components zero(non-Hermitian components $+-iu$) makes the energy spectrum(especially the real part) symmetric about zero(Figure 3(g),(h)).
	Such details of the energy spectrum have significant implications for post-quench transport dynamics, which will be discussed in the next section.

	\section{The dynamics of quench and the Physics behind}
	\subsection{Results: switching of asymmetric post quench transport}
	In the context of dynamics for non-Hermitian systems, the norm of the wavefunction may not be conserved and can become greater or less than one due to the non-unitary nature of the evolution. In quantum systems, this reflects a gain (growth) or loss in the number of particles, a concept that is widely applied in classical analogs of quantum systems \cite{ makris2008beam, el2007theory, mostafazadeh2009spectral, guo2009observation, yoshida2019exceptional}.
	We begin by considering a case where there are two non-Hermitian pairs (4 sites of alternating U and $U^{*}$) placed in a manner such that the system is PT-symmetric. Consider two sets of quenches where the initial state is an edge state localized at either of the edges ($v/w = 0.25$) to configurations $ v/w =1.125 $  and $ v/w = 1.5 $, respectively. Figure 4 illustrates the light cones formed by the probability density at each site over time. Panels (a) and (b) represent the case where \( v/w = 1.125 \), while panels (c) and (d) correspond to \( v/w = 1.5 \). One of the interesting things to note from Figure 4 (a) and (b) is the reflection in the transport from the region of PT-symmetric components, and this reflection is asymmetric for transport arising from left and right, i.e., the reflection for the wave coming from the left is lower than the reflection of the wave coming from the right. A particularly remarkable phenomenon is showcased in Figure 4(c) and (d): as the \(v/w\) ratio of the final Hamiltonian increases from $ 1.125 $ (Figure 4(a) and (b)), to $ 1.5 $, the asymmetry in reflection undergoes a striking shift. In this scenario, the reflection for the wave coming from the left becomes higher than that coming from the right. This phenomenon is especially counterintuitive when contrasted with the example of unidirectional invisibility in PT-symmetric refractive index systems \cite{lin2011unidirectional}, where a light beam interacts with a loss-gain region as in this case, in spite of no change in the non-hermitian components, the reflection asymmetry switches.\\
	\begin{figure}
		\centering
		\includegraphics[width=1.0\linewidth]{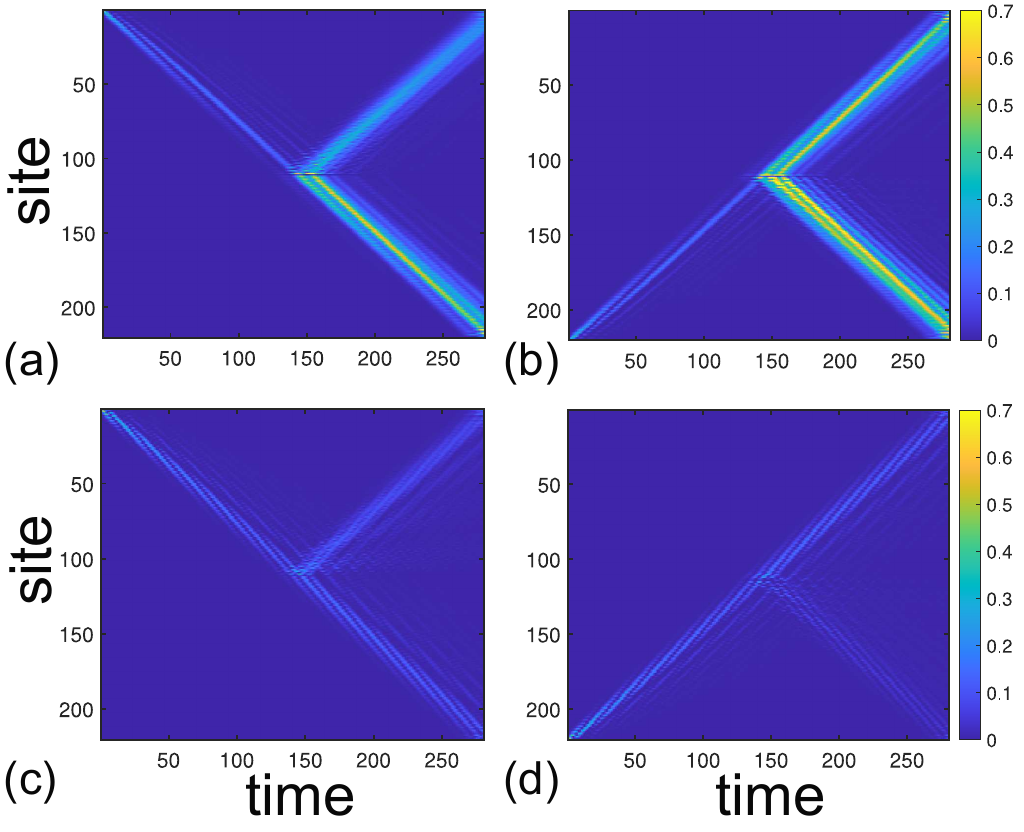}
		\caption{Light cones formed by the probability density at each site with respect to time for transport from the right(right column (a) and (c)) and from the left(left column (b) and (d)) for a quench from the initial point in parameter space, $ v/w  = 0..25 $ to a point in parameter space $ v/w = 1.125 $(top row (a) and (b)) and to $ v/w = 1.5$(bottom row (c) and (d)). The right reflection (the branching) is higher when the quench is to $ v/w = 1.125 $ and the left reflection is higher for the quench to $ v/w = 1.5$.}
		\label{fig:figure4}
	\end{figure}
	For a clearer quantitative understanding, it is useful to split the system into two halves and track the square of the wavefunction's norm within each subsystem over time, for varying values of \(v/w\) in the final Hamiltonian. Figure 5 illustrates this approach. 
	The sub-figures in Figure 5 on the left and right-hand side correspond to the initial edge state localized to the left and right-hand side of the system, respectively, and the rows correspond to different $ v/w $ values of the final Hamiltonian. It may be observed that in the left(right) columns, the blue(red) line changes its value from 1 for the initial left(right) localized edge state, and it encaptures the effect of reflection with its value being greater(less) than one signifies net gain(loss).
	Notice that for $ v/w=1.125 $ (Figure 5(a) and (b)), the changed value of the blue lines(quantifying reflection for transport from left) in Figure 5(a) is less than the changed value of the red line(quantifying reflection for transport from right) in Figure 5(b), and this scenario changes after a particular value of final Hamiltonian $ v/w $. For $ v/w = 1.25 $, the left and the right reflections are approximately the same (Figure 5 (b) and (c)), and finally, for $ v/w = 1.5 $, the reflection for transport from the left is higher than the reflection for transport from the right(the value of blue line after the kink in Figure 5(e) is higher than the value of the red line after the kink in Figure 5(f)). \\
	\begin{figure}
		\centering
		\includegraphics[width=1.0\linewidth]{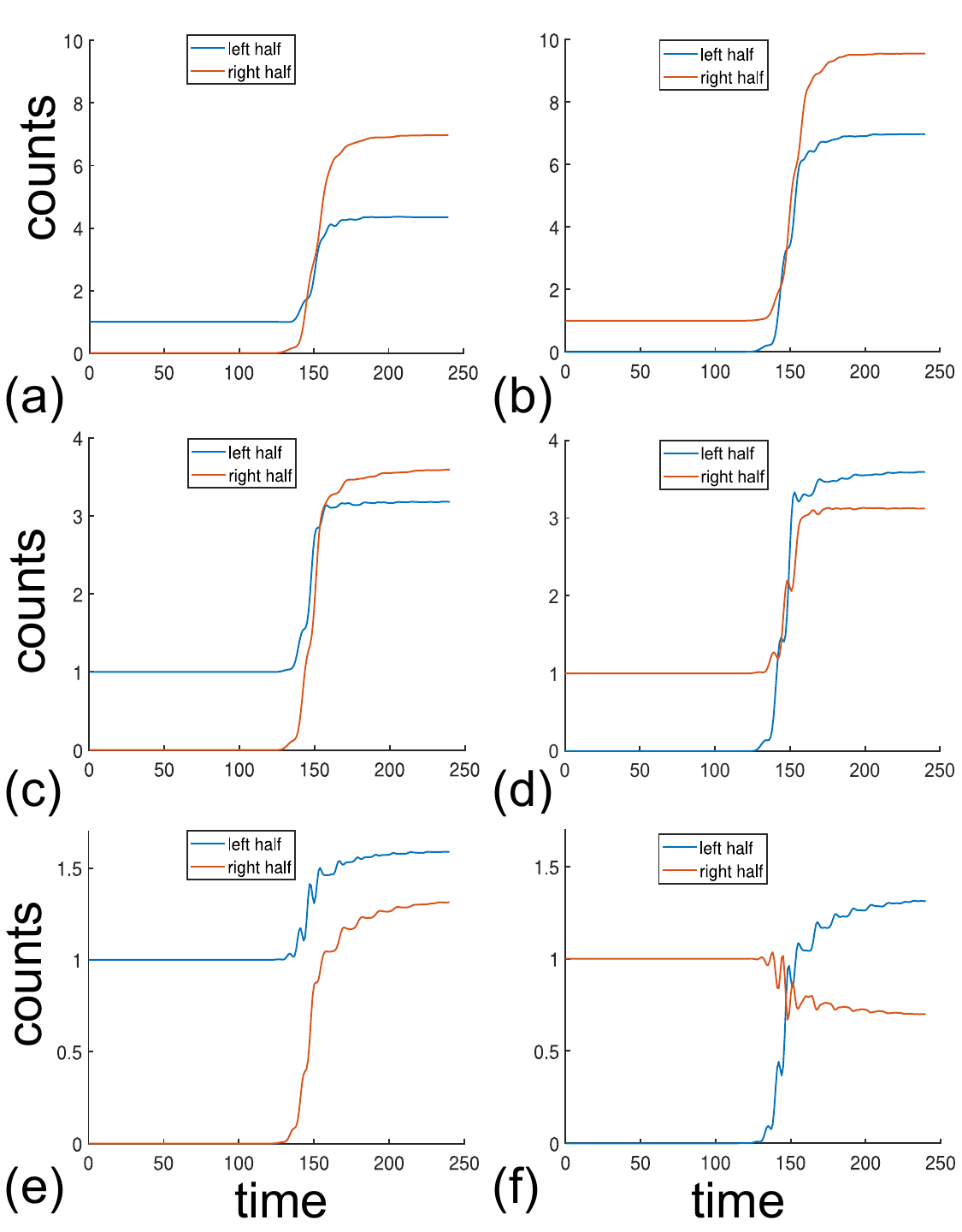}
		\caption{The square of the wavefunction's norm corresponding to the two subsystems with respect to time: the left side corresponds to the subsystem starting from site 1 to the center of the non-Hermitian pairs, and the right side corresponds to the subsystem from the center of the non-Hermitian component to the 220th site. The right column ((a),(c), and (e)) represents when the initial state is the right edge state and the left column((b), (d), and (f)) represents when the initial state is the left edge state. The top row((a) and (b)) correspond sto the case when the quench is to the point $ v/w = 1.125 $, the middle row((c) and (d)) corresponds to $  v/w = 1.25$ and the last row((e) and (f)) corresponds to the quench to $ v/w =1.5$.
		The value of 1 for the blue(red) line denotes the initially left(right) localized state and the change in this value from 1 denotes a reflection with gain or loss. For the top row, the right reflection is more, for the middle they are approximately equal and for the last row the left reflection is higher.}
		\label{fig:figure5}
	\end{figure}
	
	This fascinating physics of the switch in reflection asymmetry may be encapsulated in a single diagram, as presented in Figure 6. It depicts the ratio of the square of the wavefunction's norm corresponding to the right half for the right initialized transport to that of the left half for the left initialized transport at a chosen time($ t=240 $, see Figure 5), and this quantifies the ratio of right to left reflection amplitudes. The shift of values across $  1$ signifies a switch in asymmetry. In Figure 6, we have shown this for various configurations. Figure 6(a) captures the phenomenon of switching in reflection asymmetry for this case, and it may be noted that the ratio $ \rho_{R}/\rho_{L} $ switches from values above $ 1 $ to below at approximate $ v/w = 1.25 $. The switching in Figure 6(b), which represents a PT-symmetric system with two pairs of non-Hermitian potential $ Ur/w+-iUi/w = 0.75 \mp i 1$, is occurring for a much higher value of $ v/w $ than in Figure 6(a). Figures 6(c) and 6(d) represent the ration of right to left reflection for the systems corresponding to Figure 3 ((e),(f)) and Figure 3 ((g),(h)), and here, there is no apparent switching in the range of $ v/w $. 
	The next subsection is dedicated to discussion on the underlying physical process that led to this interesting property of switch in asymmetry.
	\begin{figure}
		\centering
		\includegraphics[width=0.8\linewidth]{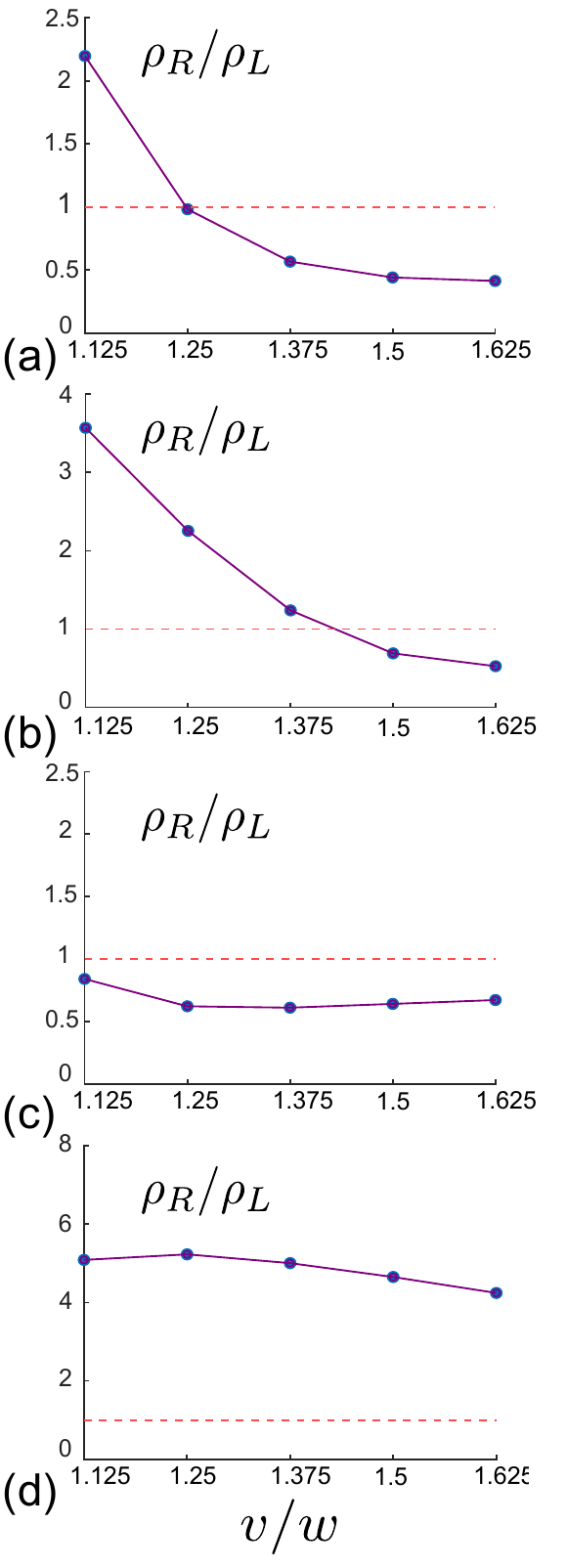}
		\caption{The ratio of the right reflection to the left reflection, as quantified by calculating the ratio of the square of the wavefunction's norm for the right subsystem(from the center of the non-Hermitian components to the 220th site) for right initialed transport to that of the left subsystem (from the $1^{st}$ site to the center of the non-Hermitian components)for left initialized transport at a specific time (t=240). The four sub-figures represent different configurations in which the non-Hermitian potential components are placed in a PT-symmetric manner on sites 109-112: (a) non-Hermitian pairs $(U_{Re}, \mp i U_{Im}) = (0.75, \mp i 0.75)$, (b) non-Hermitian pairs $(0.75, \mp i 1)$, (c) non-Hermitian pairs $(0.75, \mp i 0.25)$, and (d) non-Hermitian pairs with purely imaginary components $(0, \mp i 0.75)$.}
		\label{fig:figure6}
	\end{figure}

	\subsection{Discussions}
	To understand the post-quench transport in our hybrid system, it's important to recognize that in non-Hermitian systems, the time evolution behaves differently. The eigenvectors of the these Hamiltonians are not orthogonal, $ \bra{\psi_{n}}\ket{\psi_{m}}\neq \delta_{n,m} $, and hence does not form a complete basis, $ \sum_{n}\ket{\psi_{n}}\bra{\psi_{n}} \neq I $. However, bi-orthogonality can be defined in terms of left and right eigenvectors, i.e., for
	$ \hat{H}\ket{\psi_{n}} = E_{n} \ket{\psi_{n}} $ and $ \hat{H}^{\dagger}\ket{\phi_{n}} = E_{n}^{*}\ket{\phi_{n}} $,
	\begin{equation}
		\begin{aligned}
		    \bra{\phi_{m}}\ket{\psi_{n}}&=\delta_{m,n}\\ 
		    \sum_{n}\ket{\psi_{n}}\bra{\phi_{n}} &= \sum_{n}\ket{\phi_{n}}\bra{\psi_{n}} =I.
		\end{aligned}
	\end{equation}
Using Equation 4 in the time evolution, $ \psi(t) = e^{-i \hat{H} t}\ket{\psi_{i}} $, we obtain,
	\begin{equation}
		\psi(t)=\sum_{n}\bra{\phi_{n}}\ket{\psi_{i}}e^{-i E_{n} t}\ket{\psi_{n}}.
	\end{equation} 
	
	Here, \(\ket{\phi_{n}}\) and \(\ket{\psi_{n}}\) are the left and right eigenvectors of the Hamiltonian, respectively, while \(E_{n}\) represents the \(n\)th energy of the final Hamiltonian. Equation 5 governs the evolution of the wavefunction of the system and hence the transport which is constituted by probability density at each site with respect to time. The wavefunction amplitude at each site is the superposition of all the wavefunction $ \ket{\psi_{n}} e^{-i E_{n} t} $ weighed by the coefficients $ \bra{\phi_{n}}\ket{\psi_{i}} $. 
The wavefunctions in this non-Hermitian system exhibit spatial localization determined by the imaginary component of their energy. Specifically, they are either right- or left-centered, depending on whether the sign of the imaginary part of the corresponding energy is positive or negative.	
 The position of the special points(exceptional points for the case of PT-symmetric systems) in parameter space($ v/w$ in this case ) plays a important role as at these points there is reorganization of some of the bulk states and their energies and it is this reorganization of states and energies that leads to the switch in the reflection asymmetry.\\ 
	Figure 7 illustrates the underlying physics for the case where non-Hermitian components are placed in a PT-symmetric configuration (corresponding to Figures 3(a), (b), 4, 5, and 6(a)). The left column of Figure 7 represents the regime before the switching (\( v/w = 1.125 \)), while the right column represents the regime after the switching (\( v/w = 1.5 \)). Figures 7(a) and 7(b) plot the real versus imaginary energies. Since this is a PT-symmetric model, the imaginary energies are symmetric about zero, while the real energies are not. The color coding in the plot indicates the localization of the wavefunction: red shows that the center of mass (c.o.m. $ =\bra{\psi}x\ket{\psi} $) of the wavefunction is on the left side of the non-Hermitian pairs, green indicates the c.o.m. is on the right, and the deeper the color, the closer it is to the center of the non-Hermitian components.
	To better illustrate this wavefunction localization based on the imaginary component of their energies, Figures 7(c) and 7(d) show the imaginary part of the energies alongside the c.o.m. of the corresponding wavefunctions. As \( v/w \) increases (from the left to the right column), certain states with positive and negative imaginary values merge to zero, and the corresponding c.o.m. moves from the left and right toward the center.\\
	 The system remains in the PT-broken region for both values of \( v/w \) (as indicated by the presence of imaginary components in Figures 7((a),(c) and (b),(d)), suggesting that reflection will remain asymmetric. However, due to shifts in the energies and the corresponding localization of these states, the nature of this asymmetry changes.
	In Figures 7(e) and 7(g), two bulk states are shown as examples, clearly illustrating localization on the left (right) for negative (positive) energies. Once their imaginary energies reach zero, the states become PT-symmetric and are evenly distributed on both sides of the non-Hermitian components. Figures 7(f) and 7(h) provide examples of these symmetrized states. It is this restructuring of states(energy and spatial structure)that drives the shift in asymmetry. \\
	It is worth noting that this switching also occurs when the non-Hermitian components are arranged to break PT symmetry (as shown in Figures 3(c) and 3(d)), with the right-to-left reflection plots almost overlapping (Figure 6(a)). While the core mechanism driving the switch in transport—namely the reshuffling of states—remains the same, there is a key distinction in this case. Here, near the critical points, the energies do not reach zero but instead decrease in magnitude and reverse the spatial polarity, meaning that the c.o.m. of states with positive (negative) imaginary values shifts from right (left) to left (right).
	\begin{figure}
		\centering
		\includegraphics[width=1.0\linewidth]{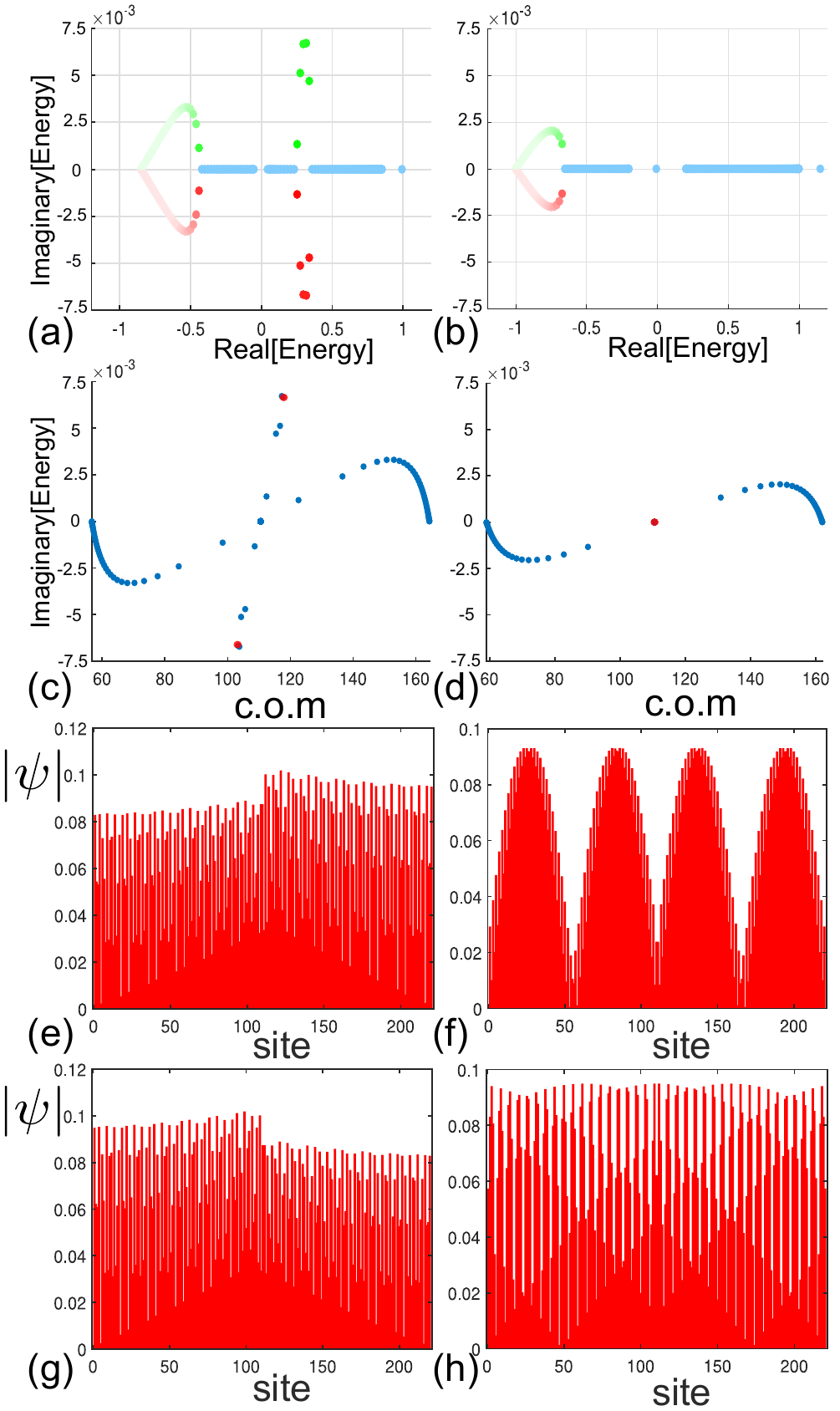}
		\caption{The reshuffling of certain bulk states in terms of energy and localization for the case when the non-Hermitian pairs $(U_{Re}/w, \mp i U_{Im}/w) = (0.75, \mp i 0.75)$ are placed in a PT-symmetric manner has been demonstrated as the $ v/w $ of the system is varied. The subfigures on the left column denote the configuration before the transition $ v/w = 1.125$ and the right column corresponds to the case after the transition $ v/w = 1.5 $. The top row((a) and (b)) is a plot of real vs imaginary components of energy, where the color coding signifies whether the center of mass of the corresponding wavefunction is on the left(red) or right(green) from the center of the pair of non-hermitian components, the deeper the color the closer it is to the center and if it is closer than a threshold (0.5), it is marked blue. As we go from the left column to the right column, certain left-centric states with negative imaginary components, as an example, (g), and certain right-centric states with positive imaginary components, as an example (e),  merge to zero simultaneously delocalizing to be symmetric about the center, as shown in (f) and (h). This physics is further conveyed through  (c) and (d) which are the plots of imaginary components of energy and the corresponding center of mass, the merging to the center may be noted. The real energies are not symmetric about zero but the imaginary energies are.}
		\label{fig:figure7}
	\end{figure}
	Figure 8 illustrates this behavior. The structure and color coding are the same as in Figure 7. The first observation is that, since the system is not PT-symmetric, the imaginary components are not exactly symmetric around zero and so are the real energies. Figures 8(a), 8(c), 8(b), and 8(d) show that as \( v/w \) varies across the switching point in transport, the imaginary components of the energy decrease in magnitude but do not vanish, while the localization of the states reverses (as indicated by the color of the energies and the c.o.m.).
	As an example, two states with positive (negative) imaginary energies are shown to be localized on the right (left) before the transition (Figures 8(e) and 8(g)). After the transition at higher \( v/w \), Figures 8(f) and 8(h) illustrate the same, now with reversed localization. Comparing Figure 8(f) with 8(e) and Figure 8(h) with 8(g) highlights this change, demonstrating the reshuffling of states in the non-PT-symmetric system. \\
	\begin{figure}
		\centering
		\includegraphics[width=1.0\linewidth]{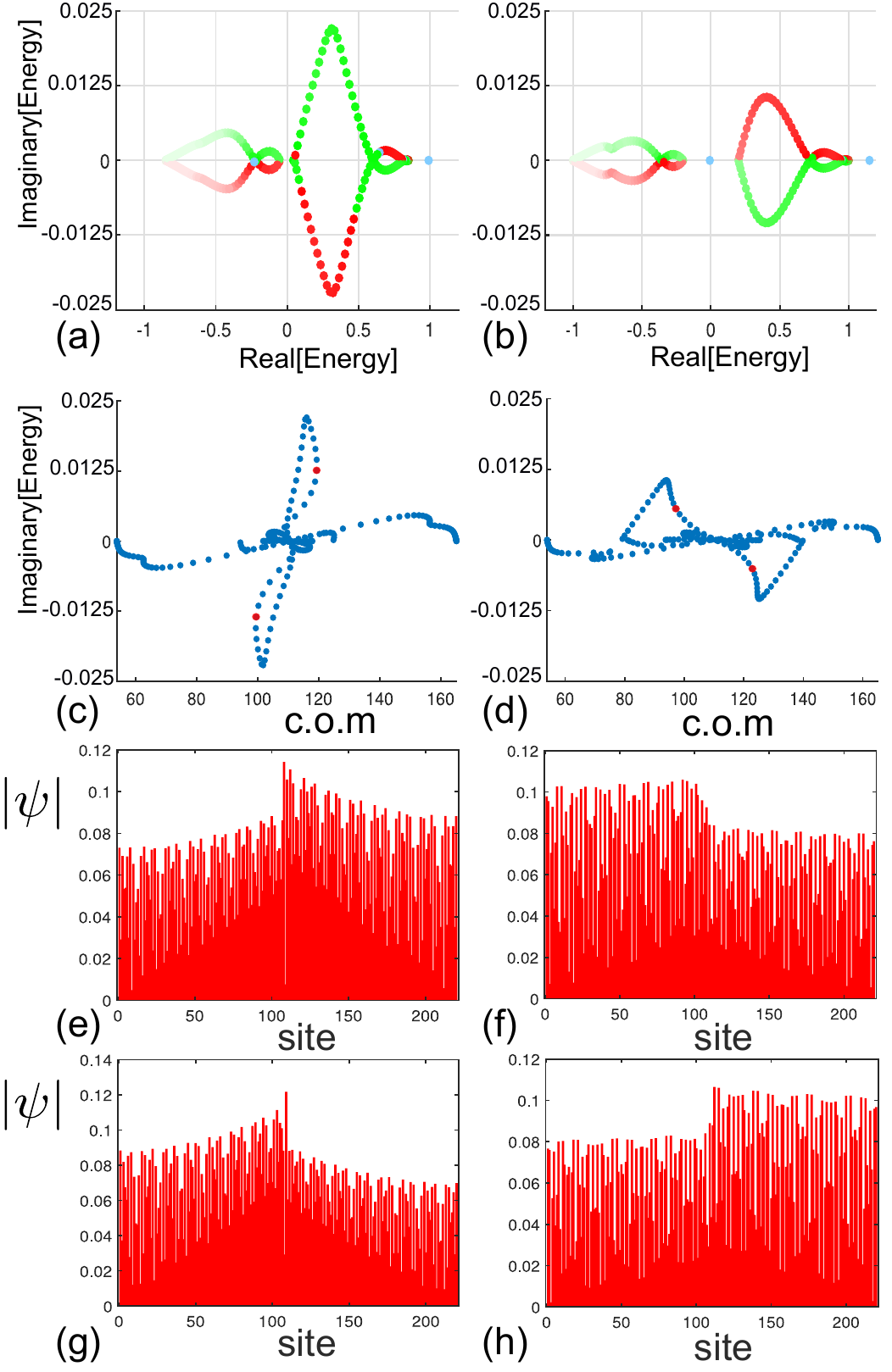}
		\caption{The reshuffling of certain bulk states in terms of energy and localization for the case when the non-Hermitian pairs $(U_{Re}/w, \mp i U_{Im}/w) = (0.75, \mp i 0.75)$ are placed in a non-PT-symmetric manner has been demonstrated as the $ v/w $ of the system is varied. The subfigures on the left column denote the configuration before the transition $ v/w = 1.125$ and the right column corresponds to the case after the transition $ v/w = 1.5 $. The top row((a) and (b)) is a plot of real vs imaginary components of energy, where the color coding signifies whether the center of mass of the corresponding wavefunction is on the left(red) or right(green) from the center of the pair of non-hermitian components, the deeper the color the closer it is to the center and if it is closer than a threshold (0.5), it is marked blue. As we go from the left column to the right column, certain left-centric states with negative imaginary components, as an example, (g), and certain right-centric states with positive imaginary components, as an example (e),  move to lower values of energy with their localization reversed, as shown in (f) and (h). This physics is further conveyed through  (c) and (d) which are the plots of imaginary components of energy and the corresponding center of mass, the reversal in their localization may be noted. The real and imaginary components of energies here are not symmetric about zero.}
		\label{fig:figure8}
	\end{figure}
	
	The location of these special points in parameter space, determined by the system's configuration, is crucial. The region of interest lies beyond the topological transition point at \( v/w = 1 \), where the survival probability of the edge states drops to zero before reemerging. The absence of switching in Figure 6(c) occurs because the transition point falls outside(at a lower value) the range of \( v/w \) values explored. In contrast, the transition in Figure 6(b) takes place at a higher \( v/w \) value than in Figure 6(a), due to the larger magnitude of the imaginary components. \\
	It’s worth highlighting that no switching occurred when the non-Hermitian components were purely imaginary (with no real component), as seen in Figure 4(d) (corresponding to Figure 3(d)). Physically, when the real part is zero, the negative imaginary component acts solely as an absorber, while the positive imaginary part functions purely as an emitter. As a result, for an incoming wave packet, the negative imaginary part on the left absorbs, while the positive imaginary part on the right emits, consistently causing the right-side reflection to exceed the left-side reflection.
	
	Interestingly, in terms of state reshuffling near the critical point, this case presents a key distinction. The energies—both real and imaginary—remain perfectly symmetric around zero throughout. The reshuffling here is highly symmetric, with imaginary values at both negative and positive real energies merging to zero, preserving this symmetry (as shown in Figures 9(a) and 9(b)).

	\begin{figure}
		\centering
		\includegraphics[width=1.0\linewidth]{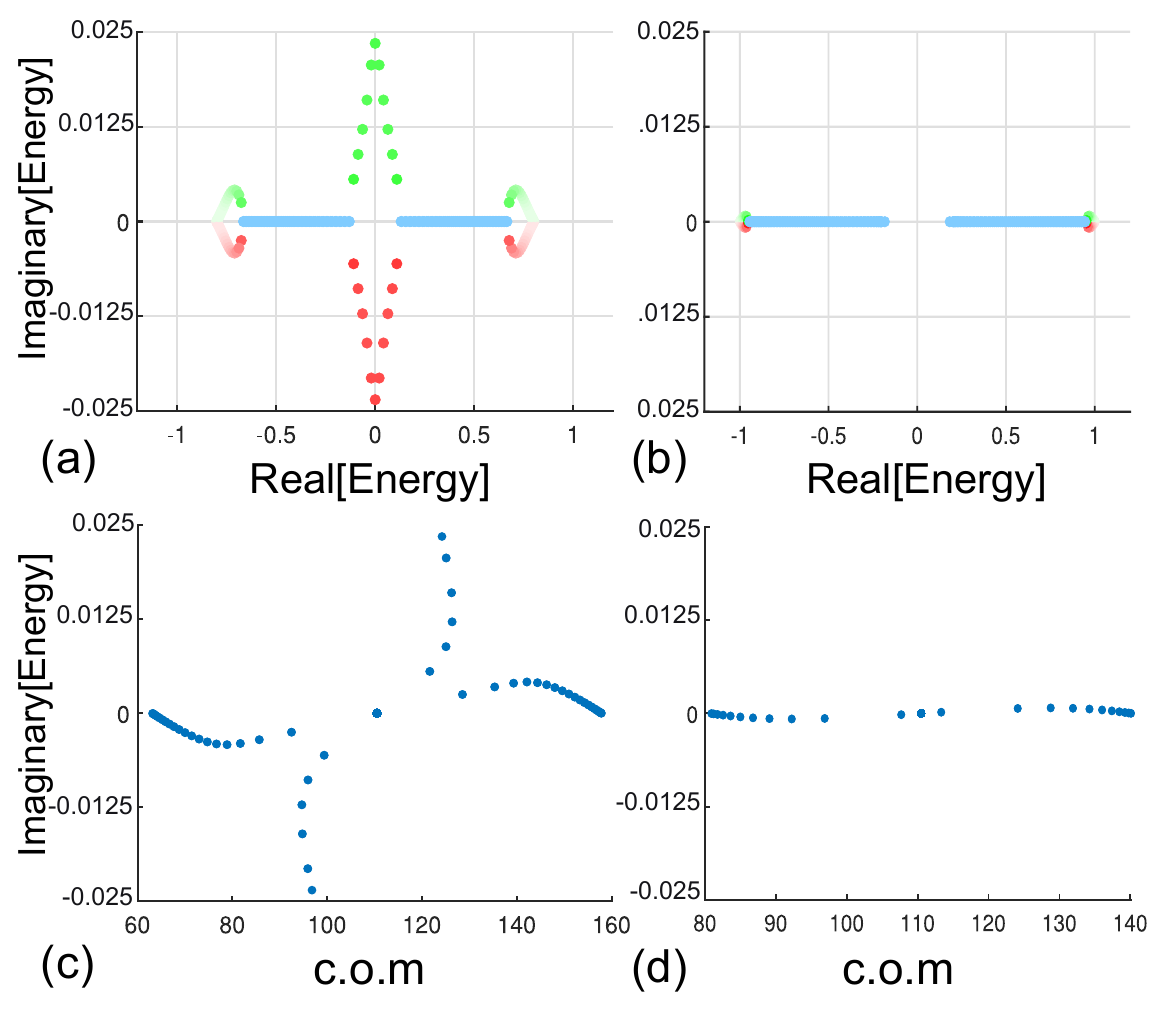}
		\caption{The symmetric nature of bulk states and the reshuffling of certain states as $v/w$ is varied is shown for the case of purely imaginary potentials $(U_{Re}/w, \mp i U_{Im}/w) = (0, \mp i 0.75)$ placed in a PT-symmetric manner. The left column represents the system before the transition at $v/w = 1.125$, and the right column shows it after the transition at $v/w = 1.5$. Panels (a) and (b) plot real vs. imaginary energy components, with color coding indicating the center of mass of the wavefunction: red for left, green for right, and deeper colors representing proximity to the center. States closer than a threshold (0.5) are marked in blue. Symmetry around zero is maintained throughout. Panels (c) and (d) further show the merging of imaginary energies to zero, plotting imaginary components vs. their corresponding center of mass.	.}
		\label{fig:figure9}
	\end{figure}

	\section{summmary}
	
	 We studied the post-quench transport in a effective model, which is non-Hermtian components embedded in a SSH like lattice. The model is chosen such that it preserves the key feature of topological phase transition, i.e., there are definite regimes in parameter space of staggered hopping amplitudes where the edge states are preserved. It is found that the transport arising as a result of quench between these regimes displays asymmetric reflection for transport arising from left and right. This asymmetry switches its polarity depending upon the regime in parameter space to which the system is quenched. We have then discussed that this switching in polarity emerges from the underlying properties of the bulk-states near the critical points, the details of which depends upon the configuration of the system. There is reshuffling of energy in terms of energy(imaginary part) dependent localization; For PT symmetric systems it is delocalization of these particular states as they transit to zero(imaginary part) and for the case of non-PT symmetric case it is the change of localization between right and left and these states don't transit to zero energy(imaginary part). The switching was not observed for any values of parameter space when the real part of the non-Hermitian components was kept zero, where the non-Hermitian components behaved as pure emitter or absorber.

	\section*{ACKNOWLEDGMENTS}
	A. Ghosh would like to express his gratitude to the University of Melbourne and the Indian Institute of Technology Kharagpur for providing a conducive research environment. Same is also thankful for the encouragement and support from his group members at both institutes, whose insights and suggestions have been instrumental in the completion of this work.
	
	\bibliographystyle{unsrtnat}
	\bibliography{References}

\end{document}